\documentstyle[prl,aps,twocolumn,epsf]{revtex}
\def\cit#1#2#3#4#5{{#1} {\bf #2}, #3 (#4)}
\def\jpsj{J.\ Phys.\ Soc.\ Jpn.\ }

\def\cras{C. R. Acad. Sci.}
\def\zncro{ZnCr$_2$O$_4$}
\def\mgcro{MgCr$_2$O$_4$}
\def\znvo{ZnV$_2$O$_4$} 
\def\mgvo{MgV$_2$O$_4$} 
\def\ymn{YMn$_2$}

\def\Bond#1#2{{\bf S}_#1\cdot{\bf S}_#2}

\def\Sp#1{{\bf S}_#1}

\def\tit#1#2#3#4#5{{#1} {\bf #2}, #3 (#4)}

\def\prl{Phys.\ Rev.\ Lett.\ }

\def\prb{Phys.\ Rev.\ B\ }

\def\jap{J.\ Appl.\ Phys.\ }

\def\jpsj{J.\ Phys.\ Soc.\ Jpn.\ }

\begin{document}

\twocolumn[\hsize\textwidth\columnwidth\hsize\csname @twocolumnfalse\endcsname

\title{Order by distortion and string modes in pyrochlore
antiferromagnets}

\author{
Oleg Tchernyshyov$^*$, R. Moessner$^\dagger$, and S. L. Sondhi$^\dagger$ 
}

\address{
$^*$School of Natural Sciences,
Institute for Advanced Study, Princeton, New Jersey 08540\\
$^\dagger$Department of Physics, Princeton University,
Princeton, New Jersey 08544
} 

\date{Version 2.0, 31 July 2001}

\maketitle

\begin{abstract}
We study the effects of magnetoelastic couplings on pyrochlore
antiferromagnets. We employ Landau theory, extending an investigation
begun by Yamashita and Ueda for the case of $S=1$, and semiclassical
analyses to argue that such couplings generate bond order via a
spin--Peierls transition. This is followed by, or concurrent with, a
transition into one of several possible low-temperature N\'eel phases,
with most simply collinear, but also coplanar or mixed spin patterns.
In a collinear N\'eel phase, a dispersionless string-like magnon mode
dominates the resulting excitation spectrum, providing a distinctive
signature of the parent geometrically frustrated state.  We comment on
the experimental situation.
\end{abstract}

]

Geometrically frustrated magnets \cite{hfmrev,lhuillier,schiffer} are
examples of strongly interacting systems: the vast degeneracy of their
classical ground states makes them highly susceptible even to small
perturbations.  By analogy with quantum Hall systems, where the Landau
levels are also macroscopically degenerate, one expects a variety of
phases in perturbed frustrated magnets, from N\'eel states to spin
glasses or liquids, with valence-bond solids along the way.

\begin{figure}
\begin{center}
\leavevmode
\epsfxsize 0.9\columnwidth
\epsffile{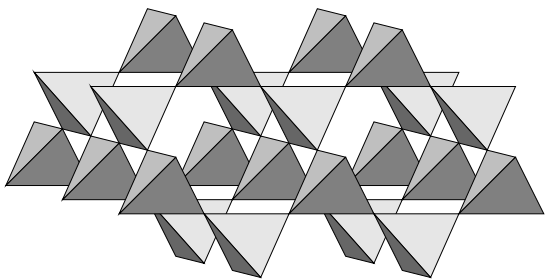}
\end{center}
\caption{The pyrochlore lattice}
\label{fig-pyrochlore}
\end{figure}

Probably the world's most frustrated spin system is the classical
Heisenberg antiferromagnet on the pyrochlore lattice
(Fig.~\ref{fig-pyrochlore}) where spins reside at vertices of
tetrahedra. The number of its classical ground states, which are
attained when total spin on each tetrahedron ${\bf S}_{\rm tot} =
\sum_{i=1}^4 \Sp{i} = 0$, is so large that, exceptionally, it does not
order at any finite temperature \cite{Chalker}.  In real compounds, {\em
deviations} from the classical Heisenberg model (e.\ g.\ dipolar
interactions, single-ion anisotropy or quantum fluctuations) determine
which ground state is selected at the lowest temperatures.

In this note, we discuss an elegant mechanism for lifting the
frustration through a coupling between spin and lattice degrees of
freedom.  The high symmetry of the pyrochlore lattice and the spin
degeneracy drive a distortion of tetrahedra via a magnetic
Jahn--Teller (``spin-Teller'') effect.  The resulting state exhibits a
reduction from cubic to tetragonal symmetry and the development of
bond order in the spin system with unequal spin correlations
$\langle\Bond{i}{j}\rangle$ on different bonds of a tetrahedron. In
the ordered phase, there are 4 strong and 2 weak bonds per
tetrahedron---or vice versa.  This phenomenon was uncovered by
Yamashita and Ueda \cite{yamashita} for pyrochlore antiferromagnets
with spins $S=1$, for which they described an AKLT-style
wavefunction \cite{aklt} with the requisite bond order.  In the
following we study this phenomenon in the semiclassical limit with
added insight from Landau theory, and discuss its consequences for the
excitation spectrum. As many of the candidate systems have moderately
large spins and order at finite temperatures, our methods should work
well---in particular, they allow us to treat the N\'eel order that can
(and experimentally does) appear in addition to the bond order.

We begin by identifying a two-component bond order parameter at the
level of a single tetrahedron and proceed to construct its Landau free
energy for states of the infinite lattice.  In one family of cases the
spin--Peierls transition can be first order, as indeed observed in many
pyrochlore
antiferromagnets \cite{plumier,niziol,ymn,mamiya,ueda-fujiwara-yasuoka,shlee}.

Once the system is in a bond-ordered phase, the lowered symmetry of
valence bonds reduces frustration of spins and sets the stage for
N\'eel (spin) ordering.  In general, we expect a separate phase
transition into an antiferromagnetic state.  However, the first-order
spin--Peierls transition may turn the system directly into an
antiferromagnet, bypassing a metastable spin--Peierls phase.  Both
possibilities seem to be realized in different
materials \cite{ueda-fujiwara-yasuoka,shlee}.
The particular type of the N\'eel order depends on the underlying bond
ordered state.  While collinear order is most easily generated, we show
that unusual coplanar states are also possible.

Finally, we obtain a striking signature of the Jahn--Teller distortion in
collinear N\'eel states.  The spectrum of spin excitations in the
distorted antiferromagnet contains a large number of modes clustered
near a finite frequency.  These magnons, a remnant of pyrochlore zero
modes, are confined to strings of parallel spins.  We therefore call
them {\em string} modes.  A resonance highly reminiscent of such
modes has been observed by S.-H. Lee {\em et al.} \cite{shlee} in
\zncro.  The strings should also be observable in \znvo\/ where a
structural distortion and a collinear N\'eel order at low temperatures
have been firmly established \cite{niziol,mamiya}.  A weaker feature
might exist in the itinerant magnet \ymn.
We turn now to the details of our results.

{\em Single tetrahedron. }  Yamashita and Ueda \cite{yamashita} have
discussed the Jahn--Teller distortion on a single tetrahedron for the
case of spins $S=1/2$.  Their analysis can be applied to general spin
virtually unchanged.  On an undistorted tetrahedron, all exchange
constants are equal and any state with total spin ${\bf S}_{\rm tot} =
0$ minimizes the classical Heisenberg energy $J\sum_{i>j}\Bond{i}{j}$.
To see which of these is selected in the presence of a small
spontaneous distortion, we minimise the sum of spin and elastic
energies.  The energy of a bond $J_{ij}\Bond{i}{j}$ depends on ionic
displacements through variation of Heisenberg exchange $J_{ij}$ with
the length of interatomic bonds and bond angles.  Expanding magnetic
and elastic energies to lowest order in the displacements $x_\alpha$
($\alpha=1\ldots12$) we obtain
\begin{equation}
E = \sum_{i,j,\alpha}(\partial J_{ij}/\partial x_\alpha)
({\bf S}_i\!\cdot\!{\bf S}_j) \, x_\alpha
+ \sum_{\alpha,\beta} k_{\alpha\beta}x_\alpha x_\beta/2.
\label{Es}
\end{equation}
As the reduction of magnetic energy is linear in $x_\alpha$, it will
beat the quadratic cost in elastic energy.  Hence a Jahn--Teller
distortion will result from the spin degeneracy.

\begin{figure}
\begin{center}
\leavevmode
\epsfxsize 0.95\columnwidth
\epsffile{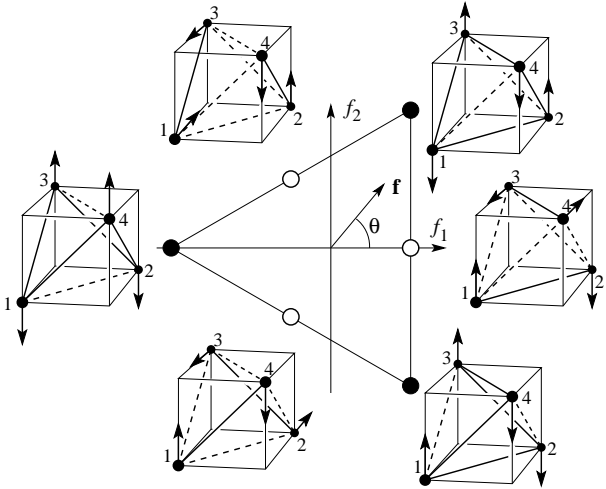}
\end{center}
\caption{
Possible values of the two components of the bond vector ${\bf f} =
(f_1,f_2) = (f\cos{\theta}, \, f\sin{\theta})$ are bounded by an
equilateral triangle in the $(f_1,f_2)$ plane.  Also shown are six
extremal spin configurations (filled and open circles).
Strong (weak) bonds are denoted by solid (dashed) lines.
}
\label{fig-triangle}
\end{figure}

For tetrahedral ($T_d$) symmetry, displacements $x_\alpha$ that
contribute to the magnetoelastic energy (\ref{Es}) must belong to the
same irreducible representation as the six bond variables
$\Bond{i}{j}$.  These bond variables can be thought of as classical
quantities or, for quantum systems, as expectation values of the
corresponding operators; our symmetry analysis applies to both cases.

In the experimentally relevant regime well below the Weiss
temperature, ${\bf S}_{\rm tot} \approx 0$, and we are left with only
three independent quantities.  One of them, the ground-state energy
$\sum_{i>j}\Bond{i}{j}$, invariant under all symmetry operations,
induces a uniform rescaling of the tetrahedron (which is not a
distortion).  The other two, $(f_1,f_2)\equiv \bf{f}$,
measure the disparity of the bond variables:
\begin{eqnarray}
 f_1 &\!=\!& 
\big[({\bf S}_1 + {\bf S}_2)\!\cdot\!({\bf S}_3 + {\bf S}_4)
 - 2 \Bond{1}{2} - 2 \Bond{3}{4}\big]/\sqrt{12},
\nonumber
\\
f_2 &\!=\!& 
(\Bond{1}{3} + \Bond{2}{4}
- \Bond{2}{3} - \Bond{1}{4})/2. 
\label{F}
\end{eqnarray}
They transform as the $E$ representation of the group $T_d$ and couple
to phonons of the same symmetry.  For a given configuration of spins,
the minimum magnetoelastic energy (\ref{Es}) can be written 
in two ways ($C_i = \rm const$):
\begin{equation}
E= -(\partial J/\partial x_E)^2 f^2/(2k_E) = 
C_0 - C_4 \sum_{i>j}(\Bond{i}{j})^2;
\label{FF}\label{C4}
\end{equation}
here $\partial J/\partial x_E$ and $k_E$ are the appropriate magnetic
and elastic constants.  Minimization of total energy is achieved when
the vector ${\bf f}$ has maximum length. From a semiclassical
analysis, we find that ${\bf f}$\ is restricted to lie in an
equilateral triangle, at the corners of which its length is maximized
(filled circles in Fig.~\ref{fig-triangle}).  This corresponds to two
weakened and four strengthened bonds, a result which is easily
rationalised in terms of classical ground states of the tetrahedron:
only one of them (modulo symmetries) fully satisfies four bonds and
completely frustrates the remaining two, which are then, respectively,
strengthened and weakened by the distortion. This state is the
collinear state, which is of course favoured by the effective
``biquadratic exchange'' (\ref{C4}).

{\em Infinite pyrochlore lattice.}  The Jahn--Teller distortion of
individual tetrahedra will analogously drive a spin--Peierls phase
transition on the infinite pyrochlore lattice. There now exist an
infinite number of phonons that could become soft.  We here discuss
the simplest but already very rich case of a phonon condensate with
lattice momentum ${\bf q} = 0$, for which all tetrahedra of a given
type distort in the same way. The two inequivalent tetrahedra (type A
and B) which make up the pyrochlore lattice reside on a bipartite
(diamond) lattice (Fig.~\ref{fig-pyrochlore}).

The symmetry group of the pyrochlore lattice, $O_h$, is enlarged from
$T_d$ by the operation of inversion through a site, which exchanges A
and B tetrahedra: $T_d \otimes C_i = O_h$.  Irreducible
representations of the cubic group $O_h$ are those of $T_d$ with an
additional quantum number, parity under the inversion.  Thus, there
are two doubly degenerate displacement modes that couple to bonds in a
nontrivial way.  The even phonon doublet $E_g$ creates a uniform
distortion of the entire lattice; the odd doublet $E_u$ causes a
staggered distortion (${\bf f}_A=-{\bf f}_B$).

\begin{figure}
\begin{center}
\leavevmode 
\epsfxsize \columnwidth 
\epsffile{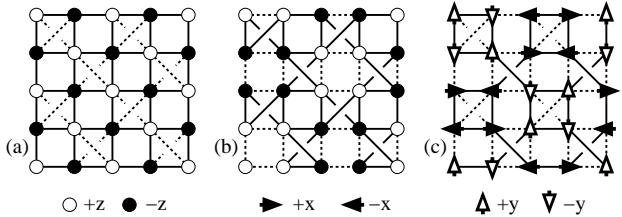}
\end{center}
\caption{(a--b) Sample collinear states of classical spins on the
pyrochlore lattice, projected along $\left<001\right>$.  
Frustrated bonds are shown as broken lines.  They
form (a) infinite chains and (b) infinite spirals.
(c) N\'eel order with collinear spins on tetrahedra $A$ and 
coplanar spins on tetrahedra $B$.  
}
\label{fig-neel}
\end{figure}

Generalising Eq.~\ref{FF} for the energy, using small subscripts
for parity and capitals for tetrahedron type, gives
\begin{equation}
E = 
-\left(\frac{\partial J}{\partial x_E}\right)^2 
\left[\frac{({\bf f}_A+{\bf f}_B)^2}{2k_g}
+\frac{({\bf f}_A-{\bf f}_B)^2}{2k_u}\right].
\label{FAFB}
\end{equation}
For $k_g= k_u$, $f_A$ and $f_B$ are separately maximized. For a
stiffer odd phonon, $k_u > k_g$, ${\bf f_A}$\ and ${\bf f_B}$\ are
in the same corner of the triangle in Fig.~\ref{fig-triangle}.  When
$k_g$ {\em slightly} exceeds $k_u$, ``antiferromagnetic'' coupling
puts ${\bf f}_{A}$ and ${\bf f}_{B}$ in two different corners.
Fig.~\ref{fig-neel}(a,b) displays spin arrangements corresponding to
these two types of bond order.  The latter has been observed in
\ymn\/ and \mgvo\/ \cite{plumier,ymn,mamiya}.

We next discuss nature of spin-Peierls phases and transitions by
considering the Landau free energy ${\cal A}({\bf g,u})$, where ${\bf
g,u}$ are the uniform and staggered bond {\em order
parameters} \cite{hbb}, ${\bf g,u} = ({\bf f}_A \pm {\bf
f}_B)/2$. In a paramagnetic state, the bond order parameters ${\bf
g,u} = 0$; note that spin--Peierls order ${\bf g,u} \neq 0$ does not
necessarily imply a N\'eel order.  The terms allowed by $O_h$ symmetry
are
\begin{eqnarray}
{\cal A}({\bf g,u}) &=& a_g g^2 + b_g g^3\cos{3\theta_g} + c_g g^4 + \ldots
\nonumber\\
&+& a_u u^2 + c_u u^4 + d_u u^6 \cos{6\theta_u} +e_u u^6 + \ldots
\label{GL}
\\
&+& b_u u^2 g \cos{(2\theta_u+\theta_g)} + \ldots
\nonumber
\end{eqnarray}
Quite generally, one can read off that the phase transition will be
discontinuous if it is driven by the even phonon $E_g$: the symmetric
cube $[E_g^3]$ contains the trivial representation making it possible
to write a cubic invariant $g_1(g_1^2-3g_2^3) = g^3 \cos{3\theta_g}$
in polar coordinates of Fig.~\ref{fig-triangle}.  On the other hand,
the transition will be continuous if it is driven mainly by the odd
phonon $E_u$.  We also note that both ${\bf g}$ and ${\bf u}$ can be
present owing to a nonlinear coupling. Then $0\neq{\bf f}_A \neq {\bf
f}_B\neq 0$, as in the case depicted in Fig.~\ref{fig-neel}(b).  Below
the transition temperature, either $a_g$ or $a_u$ becomes negative.
Omission of higher-order terms requires that $c_g, c_u > 0$.  Signs of
$b_g, b_u$, and $d_u$ will determine the nature of spin--Peierls
phases.  In our classical model, $b_u>0$, which we assume in what
follows.

\begin{figure}
\begin{center}
\leavevmode
\epsfxsize \columnwidth
\epsffile{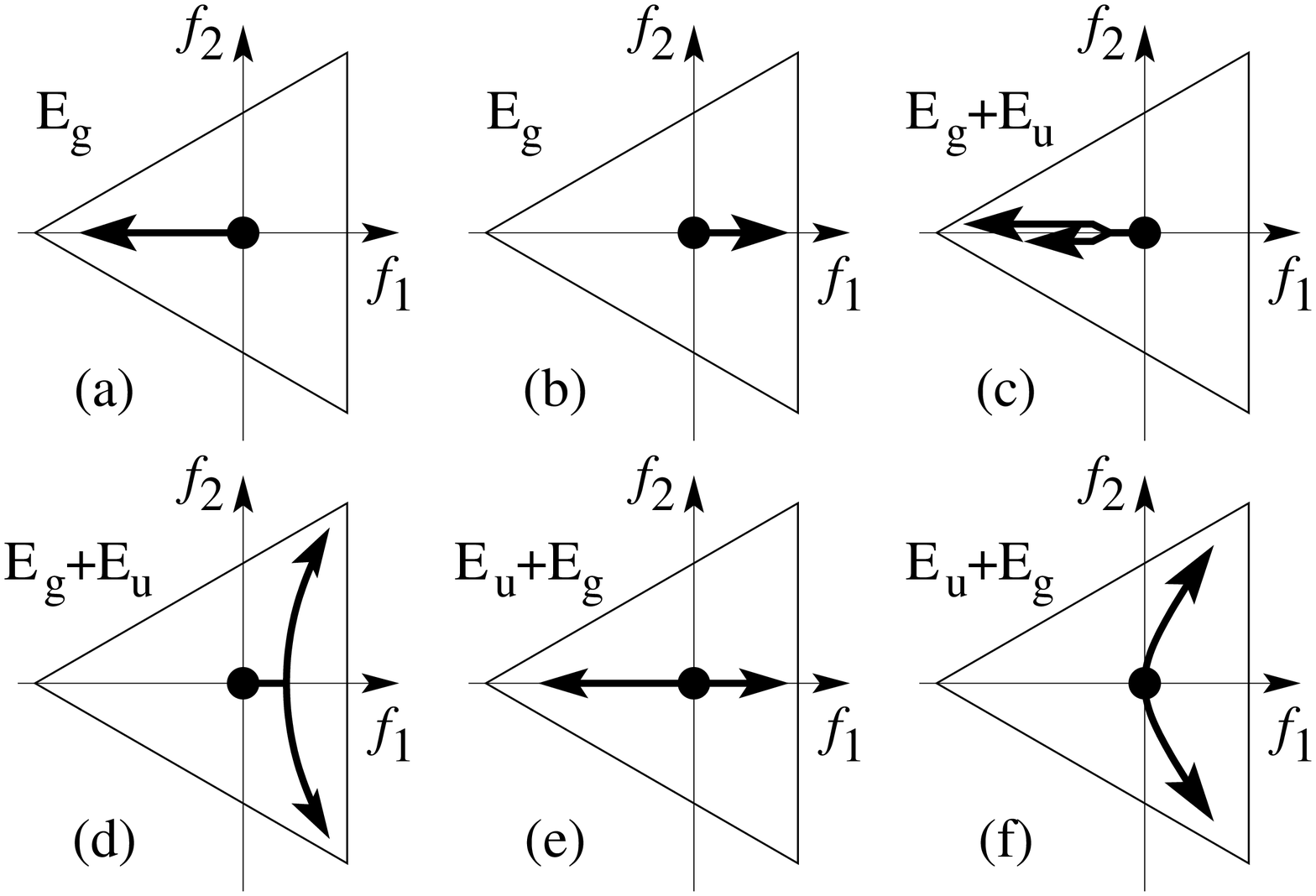}
\caption{Evolution of the spin--Peierls order parameters ${\bf f}_{A}
= {\bf g+u}$ and ${\bf f}_{B} = {\bf g-u}$.  (a--b) ${\bf g}$
condenses, ${\bf u}=0$.  (c--d) ${\bf g}$ condenses first, ${\bf u}$
appears later.  (e--f) ${\bf u}$ condenses, ${\bf g}$ is generated as
a subdominant order parameter.  In the paramagnetic phase (filled
circle), ${\bf f}_A = {\bf f}_B = 0$. }
\label{fig-VBS}
\end{center}
\end{figure}

For the phase transition driven by the uniform distortion $E_g$, only
the first line of Eq.~\ref{GL} needs to be taken into account.
Depending on the sign of $b_g$, the ordered phase features either
two strong and four weak bonds per tetrahedron
[Fig.~\ref{fig-VBS}(a)], or two weak and four strong ones
[Fig.~\ref{fig-VBS}(b)].  In the ordered phase, a growing ${\bf g}$
softens the staggered mode ${\bf u}$ through mode coupling, the third
line in Eq.~\ref{GL}.  Once $g>a_u/b_u$, there is a second
transition into a phase with both ${\bf g, u} \neq 0$ (${\bf f}_A \neq
{\bf f}_B$).  The two possible sequences of transitions are
shown in Fig.~\ref{fig-VBS}(c,d).

When the staggered phonon $E_u$ drives the transition, ${\bf g}$ can
be neglected initially, we are then left with the second line in
Eq.~\ref{GL}.  This transition is continuous (unless the ${\bf g}$
mode is nearly unstable).  The sign of $d_u$ determines the direction
of the order parameter ${\bf u}$.  The distinct possibilities are
$\theta_u=0$ (two strong bonds on tetrahedra $A$ and four strong bonds
on tetrahedra $B$) and $\theta_u=\pi/2$.  The free energy in the
ordered state with ${\bf u}\neq 0$ shows that a subdominant order
parameter ${\bf g}$ of ${\cal O}(u^2)$ arises simultaneously with
${\bf u}$ [Fig.~\ref{fig-VBS}(e,f)].  

{\em N\'eel order.} Lowering the bond symmetry relieves 
frustration and thus encourages spin ordering \cite{terao}. However,
as bond and spin ordering are not required to coincide even {\em
classically,} the former will likely precede the latter. Indeed,
frustrated systems are the ideal setting for this to happen: the
classical ground state degeneracy supplies a plethora of local
zero-energy modes \cite{Chalker}, which can allow spins to fluctuate
strongly even in the presence of bond order.  This, essentialy
classical, mechanism is distinct from the more conventional avenue of
obtaining valence-bond solids through enhanced quantum fluctuations
for small spins.  Nonetheless, if the spin--Peierls transition is
discontinuous, it can take the paramagnet directly into the N\'eel
phase, past the metastable spin--Peierls state.  Both variants seem to
occur in nature \cite{ueda-fujiwara-yasuoka,shlee}.

The phase transition from a bond-ordered state to a N\'eel phase can
also be analysed within the Landau framework by adding spin averages
$\langle {\bf S}_i \rangle$.  For the sake of simplicity, we make the
assumption that the contribution of the spin condensate $\langle
\Sp{i}\rangle \cdot \langle \Sp{j}\rangle$ will tend to reinforce,
rather than weaken or alter, bond order $\langle\Bond{i}{j}\rangle$.
Then one can use Fig.~\ref{fig-triangle} as a guide to identify the
appropriate low-temperature N\'eel phase for each of the spin--Peierls
phases of Fig.~\ref{fig-VBS}.  The spin--Peierls states shown in
Fig.~\ref{fig-VBS}(a,d) exhibit the collinear N\'eel order shown in
Fig.~\ref{fig-neel}(a,b), respectively.  The spin--Peierls phase of
Fig.~\ref{fig-VBS}(e) orders into the exotic N\'eel phase of
Fig.~\ref{fig-neel}(c) where spins on tetrahedra $A$ are collinear,
while spins on tetrahedra $B$ are coplanar. Note that, whereas the
bond order we have considered is at ${\bf q} = 0$, the concomitant
spin order need not be.

\begin{figure}
\begin{center}
\leavevmode
\epsfxsize \columnwidth
\epsffile{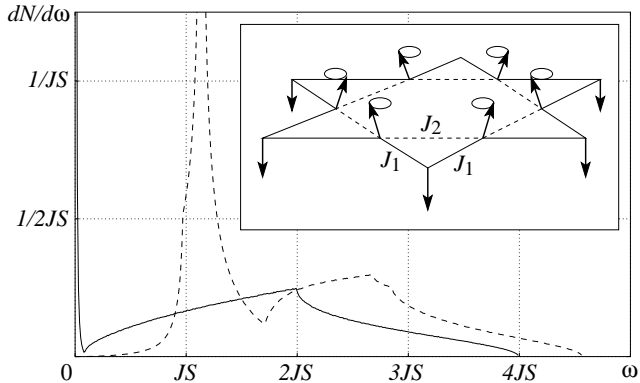}
\end{center}
\caption{Spectrum of spin waves of the elastic pyrochlore lattice for
the collinear N\'eel state of \protect Fig.~\ref{fig-neel}(a) for
$J_2:J_1 = 1:1$ (solid line) and $3:4$ (dashed line).  In the
undistorted magnet, $J_2=J_1$, half of the modes are at $\omega=0$.
The dominant peak at $\omega_0=4S(J_1-J_2)$ is due to a large number
of string modes.  Inset: Local mode on a hexagonal loop of parallel
spins.  For clarity, spins above and below the kagom\'e
plane (a [111] pyrochlore plane) are not shown.  }
\label{fig-dos}
\end{figure}

{\em String modes.} The extreme frustration of pyrochlore
antiferromagnets is reflected in their spin-wave spectrum.  A large
degeneracy of the ground state leads to a macroscopic number of
zero-frequency modes.  Although a Jahn--Teller distortion lifts the
frustration and moves the zero modes to finite frequences, we find
that the spin-wave spectrum contains a remnant of that degeneracy.
Remarkably, a large fraction of spin waves can {\em remain degenerate}
in the deformed magnet, appearing as a resonance at the frequency
$\omega_0 = 4S\delta J$.  The effective difference of exchange
couplings $\delta J$ between satisfied and frustrated bonds $J_{ij} =
J - 2 C_4(\Bond{i}{j})$ is due to the biquadratic exchange (\ref{C4}).
The resonance is a local spin wave reminiscent of the weather-vane
mode of the kagom\'e antiferromagnet \cite{weathervane}.  In the
present context, the mode is confined to {\em any} line of parallel
spins.  Such lines are abundant in collinearly ordered states: each
spin has exactly two parallel neighbors.  When successive spins on the
string precess with a phase shift of $\pi$, spins adjacent to that
line can remain unaffected because exchange field at their locations
does not precess.  (It can be shown that, despite the magnetoelastic
coupling, harmonic spin waves and phonons are decoupled in any
collinearly ordered magnet.)

For various kinds of collinear ground states, the string mode can live
on straight lines of parallel spins [Fig.~\ref{fig-neel}(a)], spirals
[Fig.~\ref{fig-neel}(b)], irregular lines, or even closed loops.  A
mode living on a short loop is truly a local resonance: the local
density of spin waves $dN/d\omega$ contains a $\delta$-function.  Open
strings are localized in only two directions and will therefore be
seen in the local density of states as a strong van Hove singularity
$dN/d\omega \propto |\omega-\omega_0|^{-1/2}$ (Fig.~\ref{fig-dos}).
The shape of the strings determines the form factor of the resonance
observable, e.g., by inelastic neutron scattering.  A recently found
resonance \cite{shlee} in \zncro\/ has the form factor of a hexagonal
string mode \cite{broholm} (the shortest possible loop,
Fig.~\ref{fig-dos}, inset).

{\em Discussion. }  We have demonstrated that a large spin degeneracy
and high symmetry of the pyrochlore network provide a robust way to
relieve spin frustration: by means of a Jahn--Teller distortion of
tetrahedra.  A transition to a spin--Peierls state with a
two-component order parameter, which can be spatially nonuniform, may
precede, or coincide with, a N\'eel transition. A unique feature of
such a distorted antiferromagnet is a large number of degenerate spin
waves confined to lines of parallel spins.  Our theory has direct
relevance to the observed magnetic and structural phase transitions
and local modes in pyrochlore antiferromagnets \ymn, \znvo, \mgvo,
\zncro, and \mgcro.
The extension of the above mechanism to other frustrated lattices
such as the triangular and kagom\'e is immediate for the case of
Ising magnets; for Heisenberg symmetry the $S=1/2$ kagom\'e system
with its high density of low energy states \cite{lhuillier}
appears to be a promising candidate for future work.

We thank G. Aeppli, C. Broholm, C. Henley, and S.-H. Lee for useful
discussions.  The work was supported in part by the DOE grant
No. DE-FG02-90ER4054442, by the NSF grant No. DMR-9978074, and by the
David and Lucille Packard Foundation.

\end{document}